\documentclass[11pt,twoside]{article}
\usepackage{asp2004}
\usepackage{psfig}
\usepackage{epsf}
\usepackage{graphics}
\usepackage{lscape}
\def\edcomment#1{\iffalse\marginpar{\raggedright\sl#1\/}\else\relax\fi}
\reversemarginpar

\begin{document}
\title{The Relation of Compact Groups of Galaxies with Larger-scale Structures}
\author{Heinz Andernach \& Roger Coziol}
\affil{Departamento de Astronom\'{\i}a, Universidad de Guanajuato,
Apartado Postal 144, 36000 Guanajuato, Gto, Mexico}

\begin{abstract}
We show that Compact Groups of Galaxies (CGs) are frequently associated
with larger-scale structures, and such associations become more
frequent for CGs selected by the same criteria at higher redshifts. Our
findings suggest that CGs are an intrinsic part of the cosmological
structure formation process and are consistent with the $\Lambda$CDM (or
``concordance'') model, but only if galaxy formation is a biased process.
\end{abstract}

\section{Introduction}

Rood \& Struble (1994; hereafter RS94) have shown that despite the
isolation criterion used to define Hickson Compact Groups (HCGs; Hickson
1982), an important fraction ($\sim$70\,\%) of HCGs are associated with
structures larger than the original groups (so-called ``loose groups''),
and a minor fraction ($\sim$5\,\%) even with rich Abell clusters of
galaxies.  In order to better understand the relation between CGs and
large-scale structures, we have extended this analysis, using different
samples of CGs, rich clusters of galaxies and poor groups encountered
at low ($z\sim0.04$) and intermediate ($z\sim0.15$) redshifts.

The organization of this article is the following. In Section~2, we
introduce the different samples of CGs, clusters and poor groups used
in our research. In Section~3, we describe briefly the method used
to establish the associations.  In Section~4, we give a summary of our
results. The results and their consequences for the large-scale structure
formation theories are discussed in Section~5. This is followed by a
brief conclusion in Section~6.


\section{The Group and Cluster Samples used here}

\subsection{Characteristics of the Samples}

In order to compare our association method with the one used by
RS94, we started out with HCGs. Confirming that our method yielded comparable
results (cf.\ Sect.~4), we extended our search to two new samples
of CGs: the {\it Southern Compact Groups} (SCGs; Iovino 2002) and the
candidate CGs from the Second Palomar Observatory Sky Survey, the
{\it Palomar Compact Groups} (PCGs; Iovino et al.\ 2003).

The sample of 121 SCGs differs from the HCGs in one subtle, but important way.
Although selected by the same criteria, the SCGs were selected
using a machine algorithm, thus eliminating the ``human'' factor.
Consequently, the SCGs are expected to form a more ``objective'' sample
than the HCGs (cf.\ Sect.~4.1). The importance of the 84 PCGs
resides in a new attempt of adapting the selection criteria used locally
to detect automatically CGs at higher redshift.

Our first rich cluster sample is that of Abell et al.\ (1989, ACO), and for their
redshift data we used the March 2004 update of the compilation of Andernach
\& Tago (1998; cf.\ also Andernach et al.\ 2004), including the full
Abell/ACO and supplementary S-catalogues.

Like for CGs, automated algorithms are now preferred over an eye selection
to detect clusters of galaxies. One such algorithm was used to derive a
catalogue of 8155 cluster candidates of the ``Northern Sky Cluster Survey''
(NSC; Gal et al.\ 2003). This forms our second list of galaxy clusters.

In order to include less rich structures like ``loose groups'', we also looked
for associations between the HCGs and the UZC-SSRS2 groups catalogue, as defined
by Ramella et al.\ (2002).

\subsection{Differences between the Abell and NSC Clusters}

One general worry associated with eye-selected samples as used in the past,
is that the ``human'' factor may have led to some elusive biases
affecting the samples. This is especially true for the Abell catalogue
compiled in the 1950's and 1980's. There are widespread opinions in
the literature, often not based on firm arguments, that the Abell sample
is increasingly biased towards higher richness clusters at higher redshifts.
Before analyzing the results of our association of CGs with clusters
of galaxies we checked the Abell and NSC samples for possible systematic
differences between them.

By correlating the NSC with the Abell clusters we found that the average
richness of NSC clusters is lower than in ACO. The mean number of galaxies
(``measured richness'' $N_{gals}$) of NSC clusters associated with an
Abell cluster is 43 (median of 41), compared to a mean of 29 galaxies
(median of 30) for those NSC clusters {\it not} associated with an
Abell cluster. A Mann-Whitney test performed on the data confirms the
difference between the median richness at the 95\% confidence level.
(Non-parametric tests were used because the two samples were found to
have non-Gaussian distributions of $N_{gals}$ and $z$.)

By comparing their distributions in redshift, we also found the NSC
to reach slightly higher redshifts than the Abell sample. The means
are $z=0.1541$ (median $z=0.1468$) for the NSC associated to an Abell
cluster and $z=0.1616$ (median $z=0.1565$) for the non-Abell-associated
NSCs. This difference was also confirmed at a 95\% confidence level using
a Mann-Whitney test. Note, however, that the NSC clusters generally have
estimated redshifts only.

To estimate the amount of incompleteness of the samples in terms of
richness at different redshift, we divided the two samples in redshift
bins of width $\Delta\,z=0.05$, centred at $z=0.5, 0.1, 1.5, \ldots$,
and into different richness classes.  For the NSC cluster richness
we used $N_{gals}$, the number of galaxies in each cluster, and split
them into five bins, roughly consistent with the definition of the
Abell richness class.  We compared the observed frequency of clusters of
different richness with that expected from a linear increase in volume
at higher~$z$ (assuming no cosmology).  For both samples (Abell and NSC)
we found no evidence of incompleteness out to $z\sim0.2$ for rich clusters
($R>0$ for ACO; $N_{gals}\ge30$ for NSC).  However, both samples show a
rapid decline of poor clusters ($R=0$ for ACO and $N_{gals}<30$ for NSC)
at $z>0.05$ for the NSC and for $z>0.1$ for the ACO sample.

 From our comparison we conclude that for comparable richness the two samples
show a similar distribution in redshift. Contrary to the general belief,
both samples show the same level of completeness up to $z=0.2$, consistent
with results found by Miller \& Batuski (2001) for the ACO sample. The
only obvious bias encountered in the Abell sample is that against poor
clusters of galaxies, which are more numerous in the NSC sample. Quite
enigmatic, however, is the rapid decrease with $z$ for these smaller-mass
systems in both samples, despite the (confirmed) greater depth of the
NSC. In fact, the decrease of the fraction of poor clusters is even more
pronounced in the NSC than in the ACO sample, and it occurs at lower $z$.
It is as if poor structures ``flourished'' only at low $z$.

\section{Search Method}

\subsection{Association Algorithm}

Our association algorithm involves two steps. First, for each group in the
samples we determine if its centre fell within one Abell radius projected
distance of a cluster. The Abell radius is defined as $1.5\,h^{-1}$Mpc
($H_{0}=100\,h^{-1}\,$km\,s$^{-1}$\,Mpc$^{-1}$, Abell 1958), which
corresponds to an apparent radius of 1.7\,$'/z$, where $z$ is the cluster
redshift. Second, we verified that the associations found are physically
real. This is done by visually inspecting the group and cluster
environment on digitized sky surveys using Aladin (aladin.u-strasbg.fr)
within the Astrophysical Virtual Observatory (AVO; www.euro-vo.org),
and by comparing the redshifts of the CGs and clusters. For the redshift
of the HCGs and SCGs we used the mean redshift of the galaxies included
in NED (nedwww.ipac.caltech.edu).  For the PCGs, a photometric redshift
was estimated by applying the redshift-magnitude formula in Gal et al.\
(2003) to the mean colors and magnitudes of the PCG galaxies, as given
by Iovino et al.\ (2003).

For the association of CGs with poor structures we used an algorithm
offered at the Strasbourg Astronomical Data Centre (CDS; vizier.u-strasbg.fr)
that allows to cross-correlate catalogues.
For the HCG vs.\ UZC-SSRS2, all UZC-SSRS2 groups within 25\,$'$
projected distance from an HCG group centre were pre-selected.
For the PCG vs.\ NSC, given their typically higher redshift, all NSCs within
15\,$'$ projected distance from a PCG centre were pre-selected as
candidate pairs.  Like for the other associations, each pair was then
inspected visually, and pairs with discordant redshifts were discarded.

\subsection{Association Types and Bautz-Morgan Classes}

During our visual inspection it proved important to distinguish three
``association types'' depending on the kind of structure a CG seemed to
form within the associated larger-scale structure. In numerous cases,
the CG seemed to form a sort of substructure of the associated larger
scale system (SS type). Very naively, we would have thought that all
associations of CGs with clusters would have been of SS type. To our
surprise, however, we found that not few CGs, especially SCGs, formed the
most luminous, often central, galaxies in a cluster (ML type).  A third
type of association was found necessary to describe associations of CGs

\begin{figure}[!t]
\psfig{file=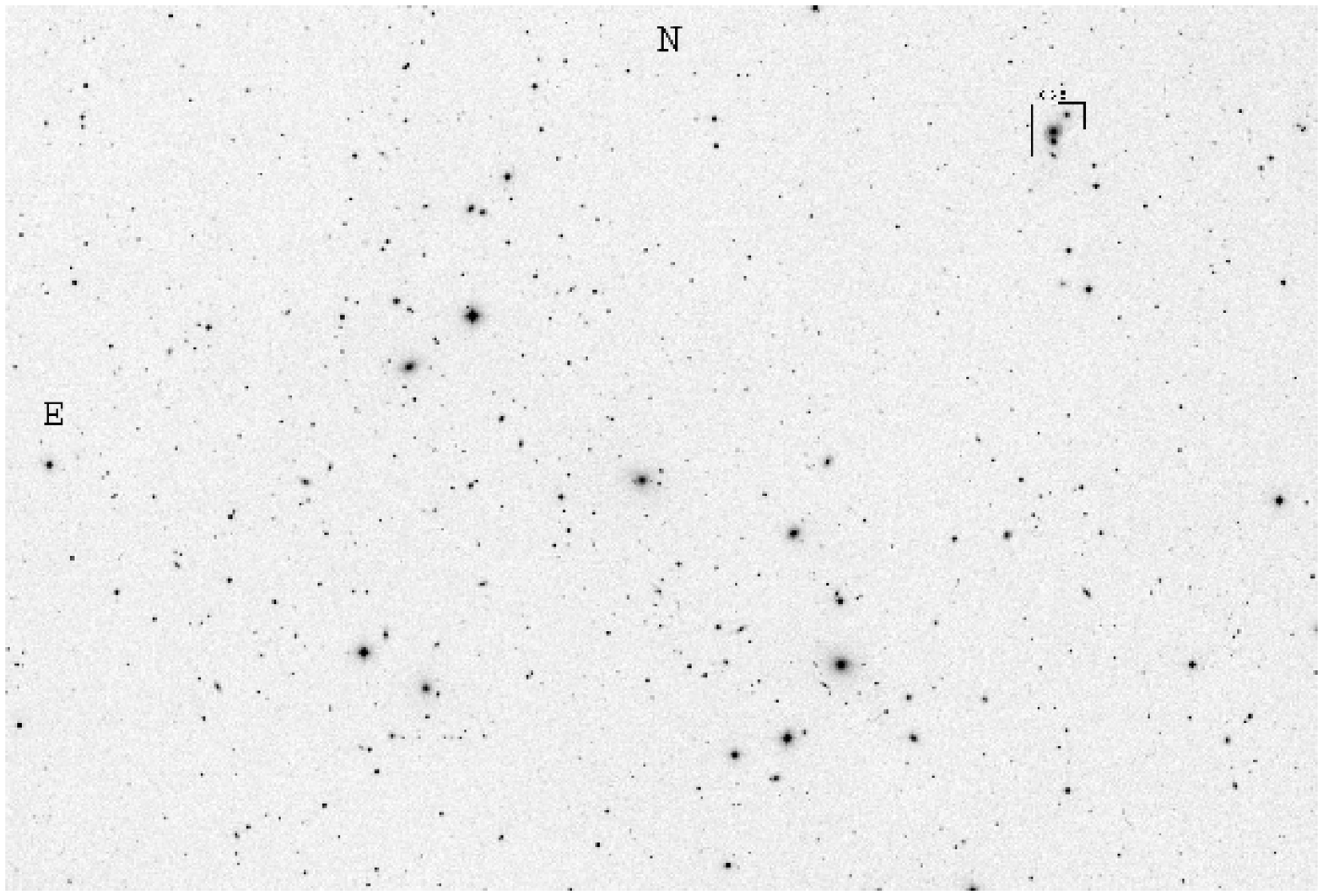,width=13.4cm,bbllx=31pt,bblly=82pt,bburx=582pt,bbury=400pt,clip=}
\caption{Example for association type ``SS'' (SubStructure) at $z$=0.041.
HCG\,5 consists of the three brightest galaxies inside the box of 2$'\times\,2'$
at upper right.  The centre of the Abell cluster A~76 is indicated by
a gapped plus sign below centre. Its Abell radius is 41\,$'$.
The image covers 47$'\times\,26.6'$. }
\label{hcg5}
\end{figure}


\begin{figure}[!h]
\psfig{file=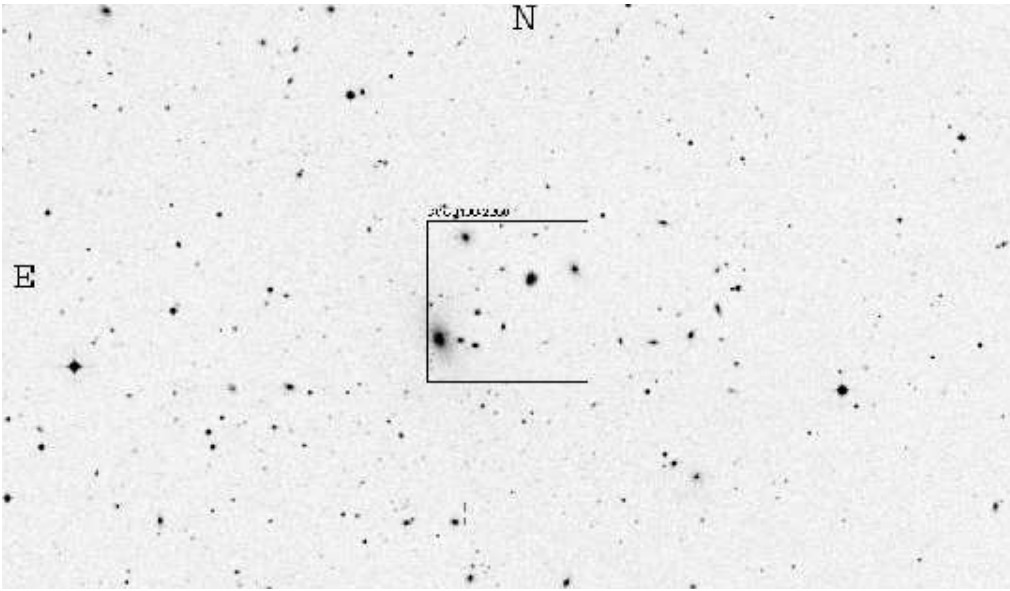,width=13.4cm,bbllx=32pt,bblly=32pt,bburx=572pt,bbury=346pt,clip=}
\caption{Example for association type ``ML'' (Most Luminous structure).
SCG~0100$-$2208 consists of the four brightest galaxies inside the box of
4$'\times\,4'$, including the central cD of Abell cluster A\,133.
The cluster centre is indicated by a gapped plus sign $\sim4'$ south
of the cD. Its Abell radius is 30\,$'$.  Both systems are at $z$=0.056.
The image covers 25.3$'\times\,14.6'$. }
\label{scg}
\end{figure}

\begin{figure}[!t]
\psfig{file=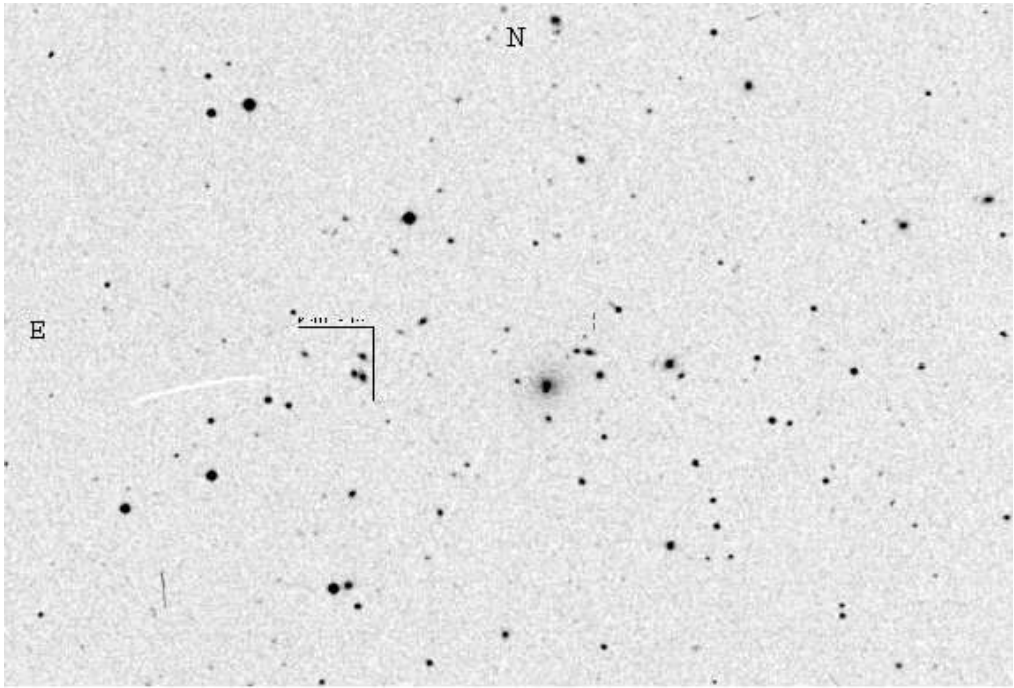,width=13.4cm}
\caption{Example for an association type ``AML'' (Among the Most Luminous).
PCG\,J1109+4133 consists of the four galaxies inside the box of 1.2$'\times\,1.2'$.
The centre of the Abell cluster A1173 is indicated by a gapped plus sign
$\sim1'$ NW of the brightest galaxy at centre.  Its Abell radius is 22\,$'$.
Both systems are at $z$=0.076. The image covers 16$'\times\,11'$.}
\label{pcg}
\end{figure}

\noindent
with clusters at higher redshifts. In this case, the CG formed a
subgroup of the most luminous galaxies in a cluster without dominant
galaxies. We dubbed this as the ``among the most luminous'' (AML) type.
A typical example of each association type is shown in Figs.\ 1--3,
and the possible physical meanings of these different association types
will be discussed further in Section~4. In particular, some level of
correlation will be shown to exist between the association types, the
richness, and the Bautz-Morgan (BM) types (Bautz \& Morgan 1970) of the
Abell clusters. For simplicity, we merged the BM types (taken from ACO)
into two classes: the ``early-type'' includes BM~I, I-II and II, and the
``late-type'' includes BM~II-III and III.

\section{Results}

\subsection{Association of CGs with Abell Clusters}

The results for the association of the three samples of CGs with Abell clusters
of galaxies is summarized in Table~1.

\begin{table}[!ht]
\begin{center}
\caption{Association of CGs with Abell clusters of galaxies}
\smallskip
{\small
\begin{tabular}{lrrcrrcrrccrr}
\tableline \noalign{\smallskip}
Sample & N &\% & $~~~\langle z\rangle$& \multicolumn{3}{c}{~~~Association type~~~}&
\multicolumn{4}{c}{Abell~Richness~Cl.}& \multicolumn{2}{c}{BM-type}\\
       &   &   &   & ~SS & ~ML & AML & 0 & ~1 & ~2 & 3 & early & late\\
\noalign{\smallskip} \tableline \noalign{\smallskip}
HCG   & 6 &6 & 0.04& 5 & 1&  &5 &  &  &1 &  2~~ & 3~~ \\
SCG   & 21&17& 0.04& 9 &12&  &18&2 &1 &  & 14~~ & 5~~ \\
PCG   & 23&27& 0.12& 10& 5&9 &3 &15&4 &2 &  5~~ &18~~ \\
\noalign{\smallskip} \tableline
\end{tabular}
}
\end{center}
\end{table}

The number of HCGs associated with Abell clusters is consistent with
what RS94 found, suggesting our method to be comparable.  However,
the frequency of Abell-cluster associations turns out to be three times
higher for the SCGs. \linebreak[4] Table~1 shows that this is {\it not}
due to a redshift difference between HCGs and SCGs. We rather seek the
explanation for this in the only difference between these samples, i.e.\
in their method of selection (automated vs.\ eye selection). Examining
the association types, it seems that the higher fraction of SCGs
associated with Abell clusters is dominated by the higher number of ML
types encountered in the SCGs.  This, in turn, would rather be expected
based on the different selection methods: any researcher selecting
CGs by eye would have discarded CGs of association type ML for being
non-isolated, even if they satisfied the formal CG selection criteria,
while these objects were kept in the automated selection of the SCGs.
This result appears to confirm that an automated selection yields a more
objective sample of CGs, and, consequently, of their possible associations
with large-scale structures.

Table~1 shows that the PCG-ACO associations occur on average at three
times higher redshift than the SCG-ACO or HCG-ACO ones, confirming
the aim of the PCG authors to find CGs at higher redshift.  However,
an interesting change in the distribution of association types occurs
for the PCGs as compared to SCGs, with the occurence of the AML type,
which seems to appear only in clusters associated with the PCGs, and a
shift away from the ML type.

A notable difference is also seen for the richness and BM types of the
associated clusters found at higher redshift: the richness increases and
the BM type changes from nearly all early-type at lower $z$ to nearly
all late-type at higher $z$. It is noteworthy that these changes seem
to correlate well with the inclusion of the AML type. Indeed, with
the richness increasing we should naturally expect a higher number of
luminous galaxies in clusters at higher redshift. The later BM type of
these clusters, on the other hand, suggests that none of these luminous
galaxies dominate over the others. This is consistent with the AML type,
which suggests that some of these luminous galaxies in these clusters
form structures similar to CGs.

The consistency of the above results suggests that the SCGs and PCGs
are formed of similar structures at low and high redshift, as expected
based on the similar method applied to detect them. What really
differs, consequently, is the nature of the clusters with which they are
associated. These structures seem to become richer and of later BM type
at higher redshift.

The variation of richness and BM type of clusters associated with CGs
at intermediate redshift is a significant phenomenon. Comparing the frequency
of clusters with different richness at different redshift, from $z=0.05$ to
$z=0.1$ the fraction of $R=0$ clusters decreases only by a factor of 1.3, while
we observe a decrease by a factor of $\sim$6 passing from the ACO-SCGs to the
ACO-PCGs. The fraction of ACO clusters with $R\ge1$ increases only by a
factor of $\sim$2 from $z\sim0.05$ to 0.1, while we observe an increase by a
factor of $\sim$5 passing from the ACO-SCGs to the ACO-PCGs.
For the BM types, at $z \sim 0.1$ the late types constitute 52\% of all ACO
clusters, compared to 48\% for the early BM types (BM$\le$II).
There is consequently a 50-50\% chance to find an early or
late BM type at this redshift in the ACO catalogue. This is in contrast with
the 78\% late-type BMs of the clusters associated with PCGs. In fact,
no BM I or I-II cluster is present in the ACO-PCG pairs.

We conclude that the increase in richness and the preference of late
BM types for ACO clusters associated with PCGs at intermediate redshift
cannot be explained by an incompleteness or any other selection effect
in the Abell catalogue.

\subsection{Association of CGs with poorer Structures}

Our results for associations of CGs with poorer clusters (NSC clusters
{\it not} \linebreak[4]
associated with an Abell cluster) and loose groups (UZC-SSRS2) are shown in
Table~2.

\begin{table}[!ht]
\begin{center}
\caption{Association of CGs with poorer structures}
\smallskip
{\small
\begin{tabular}{lccccccc} \tableline \noalign{\smallskip}
Samples & N &\% & $\langle z\rangle$& $\langle N_{gal}\rangle$&
\multicolumn{3}{c}{Association type}\\
       &   &   &                    & & SS & ML & AML \\
\noalign{\smallskip} \tableline
\noalign{\smallskip}
HCG vs UZC     & 16 & 16\,~    & 0.04 & 10 &    &   15  &  1\\
PCG vs NSC     & 21 & 25\,~    & 0.10 & 28 &  1  & 10 &  10\\
\noalign{\smallskip} \tableline
\end{tabular} }
\end{center}
\end{table}

A total of 33\% of the HCGs are associated with UZC-SSRS2 groups. However,
not all of these associations imply a larger-scale structure. In their analysis
of the UZC-SSRS2 groups, Ramella et al.\ (2003) distinguished between poor and
rich groups, the latter having $N_{gal}\geq5$ member galaxies. Making this
distinction we find that 16\% of the HCGs are associated with rich groups,
having 10 galaxies on average. The mean redshift of these groups is $z=0.04$.

We found that 25\% of PCGs are associated with an NSC cluster that
is {\it not} part of the Abell sample.  The mean number of galaxies in
these clusters is 28. This is significantly lower than the mean number of
galaxies (47) for those NSC associated with an Abell cluster. The mean
redshift of the non-associated NSC clusters is $z=0.1$. No difference in
redshift can be distinguished between the non-Abell-associated and
Abell-associated NSCs. The NSC thus maps lower-richness clusters than
ACO at the same intermediate redshift.

\subsection{Association of CGs with Large-scale Structures}

Taken as a whole, we find that CGs are frequently associated with larger-scale
structures, confirming the conclusion reached by RS94. For the HCGs the
total fraction is 20\%. However, considering that this sample was selected by eye,
and consequently, is biased against the ML association type, the real fraction
of associations with larger-scale structure could be much higher.
This is confirmed by the results obtained for the SCGs: 17\% are associated
with Abell clusters, almost three times more than for the HCGs. As we go
to higher redshift (from $z=0.04$ to $z=0.1$), the association of CGs
with large-scale structures becomes much more evident. The total fraction of
PCGs associated with larger-scale structures (ACO$+$NSC) reaches 52\%.

Now, if we consider poorer structures like loose groups, as we have done
comparing the HCG with the UZC-SSRS2 groups, a fraction higher by a factor of
2--3 may be the general rule for associations of CGs with larger-scale structures.
This could yield possible association fractions as high as 40-60\% for local
($z\sim0.04$) CGs and possibly 100\% at higher redshift ($z\ga0.1$), assuming
that structures like loose groups exist at these redshifts.

\section{Discussion}

Considering the isolation criteria used to define CGs, it may seem
surprising to find CGs associated so frequently with larger-scale
structures. From the point of view of structure formation theories, however,
this result is, in fact, quite natural. Within this context, the isolation
criterion used to define CGs only allows mapping a particular level of galaxy
density (or mass density) in a universal structure formed mostly by dark
matter. Within this context, therefore, CGs associated with larger-scale
structure are not only expected but in a sense are an inescapable phenomenon.

Moreover, by studying the kind of association CGs form with larger-scale
structures it may be possible to shed some light on the formation process
itself. For example, West (1989) found that groups of galaxies exhibit a
tendency to be aligned with their neighbors on scales of $\sim15-30\,h^{-1}$\,Mpc.
This led him to suggest that groups formed after superclusters,
and that consequently, massive structures developed before smaller-mass ones.
Similarly, Einasto et al.\ (2003) found that loose groups in the
neighborhood of rich clusters are more massive and luminous than groups on
average. They suggest that this phenomenon indicates that the formation
processes of loose groups were somehow influenced by the more massive
structures, which, once again, seems to imply that massive structures
formed before smaller-mass ones. Recently, Coziol, Brinks \& Bravo Alfaro
(2004) found that the galaxies in HCGs that are associated with massive
structures are more evolved morphologically and less active, than those
in CGs related to smaller-mass systems. In terms of structure formation,
such an observation could also be interpreted as evidence that massive
structures formed before less massive ones.

In view of the above results, it is highly significant to find a stronger
association of CGs with large-scale structures at higher redshift.
Taken at face value, it suggests that the structures with which
CGs are associated at higher redshift are richer and consequently more
massive (Girardi et al.\ 1998) than those at lower redshift. This would
be expected if massive structures formed before less massive ones.

The question remains, however, whether our interpretation of the
association of CGs with large-scale structure is legitimate. Consider the
opposite point of view, i.e.\ that galaxies in the Universe form systems
of different sizes that have no relation with each other. By definition,
therefore, different isolation criteria would lead to independent isolated
systems similar to either CGs or clusters of galaxies.
Adopting this point of view, however, would oblige us to reject arbitrarily
a large fraction of the CGs now detected (since they are found to be part
of a larger-scale structure, {\it despite} the fact they conform to the
same isolation criteria). If our estimate based on the comparison of
the HCGs with the UZC-SSRS2 is correct, a conservative attitude would
then be to consider as suspicious about half of all CGs found at
$z\le0.05$, and even the vast majority at higher redshifts.

Note that from the point of view of cosmology, adopting the extreme
position as above stated would not change our conclusion. It could imply
that CGs simply do not exist at higher redshift, which would be consistent
with a view that massive structures formed first, and less massive
ones, like CGs, appear only recently (at low $z$). Even adopting a
less stringent position, accepting for instance that CGs may have a
SS relation with clusters, does not change the conclusion, because the
remaining structures would still seem to become richer and consequently
more massive at higher redshifts.

In structure formation theories, there are two main streams of ideas. The
first is the top-down scenario, which is based on hot dark matter and
which predicts that large-scale structures should form first, while
smaller structures fragment from it at a much later time. The second is
the bottom-up scenario, which is based on cold dark matter and which
predicts that large-scale structures form by the gradual and continuing
mergers of smaller-mass systems.  Our observations on CG associations
with large-scale structures are consistent with the view that
large-scale structures formed before smaller-scale ones. It is important
here to emphasize that our results constitute the fourth evidence based
on groups of galaxies in favor of this scenario, the three previous ones
being the studies by West (1989), Einasto et al.\ (2003), and Coziol,
Brinks \& Bravo Alfaro (2004). Consequently, this particular result seems
now well established. But does it automatically support the top-down
scenario (as suggested by West 1989)? Davis et al.\ (1985) have shown
that if biased galaxy formation is included in the bottom-up model, i.e.\
galaxies forming first and preferentially in high-density regions, then
formation of small structures could be delayed significantly in favor
of larger-scale ones. The same bias was obtained naturally by Benson
et al.\ (2001) for a $\Lambda$CDM model. Our findings could thus be
consistent with the $\Lambda$CDM (or ``concordance'') model if biased
galaxy formation exists.

\section{Conclusion}

Using different samples of CGs and clusters of galaxies we confirm the results
previously obtained by RS94, which is that CGs are frequently associated with
large-scale structures. There are two ways to interpret this result. One is
to view this high frequency of association as an evidence of a failure of the
method by which CGs are selected. Adopting this point of view, we would thus be
obliged to reject as much as 40-60\% of the CGs found locally and nearly all
of those found at intermediate redshift. Hence, it is doubtful whether the
remaining CGs would play any role in the formation of large-scale structure
at any redshift.  The alternative is to view the isolation
criterion as a relative one: it only allows mapping a particular level of
galaxy density (or mass density) in a universal structure with a continuous
variety of scale sizes, and formed mostly by dark matter.
According to this interpretation, the formation of CGs is part of the
cosmological structure formation process and their associations with
large-scale structures is only natural.

Adopting this last point of view, we found that CGs at higher redshift
seem to be associated with more massive structures, suggesting that
massive structures developed before less massive ones. To our knowledge
this is the fourth evidence based on galaxy group studies in favor of
such a scenario (the other three are West 1989; Einasto et al.\ 2003;
Coziol, Brinks \& Bravo Alfaro 2004). This phenomenon seems therefore
well established. Assuming the bottom-up scenario is correct, the evidence
would thus suggest that the process of galaxy formation is a biased one:
galaxies formed first and preferentially in high-density regions.

If the latter interpretation is correct, then, independent of the cosmology,
any galaxy structure that can be observed at high redshift needs to be among
the most massive ones.

\acknowledgments This research has made use of the Aladin software
and the Vizier server, developed at CDS, Strasbourg, France.  We used
the NASA/IPAC Extragalactic Database (NED) which is operated by the Jet
Propulsion Laboratory, California Institute of Technology, under contract
with the National Aeronautics and Space Administration. HA acknowledges
financial support from CONACyT grant 40094-F to attend this meeting.
RC is supported by CONACyT grant 40194-F.

\end{document}